# Reactive Transport Simulation of Silicate-Rich Shale Rocks

# when Exposed to $CO_2$ Saturated Brine Under High Pressure and High Temperature

Shaziya A. Banu[1], Venkata R.S.B. Varanasi[2], Arash Noshadravan, Ph.D.[1], and Sara Abedi, Ph.D.[*1]

[1]Texas A&M University, College Station, TX-77840, USA.

[2]Occidental Petroleum Corporation, Houston, Texas, USA.

[*]sara.abedi@tamu.edu

**ABSTRACT:** This study examines the feasibility of carbon dioxide storage in shale rocks and the reliability of reactive transport models in achieving accurate replication of the chemo-mechanical interactions and transport processes transpiring in these rocks when subjected to $CO_2$ saturated brine. Owing to the heterogeneity of rocks, experimental testing for adequate deductions and findings, could be an expensive and time-intensive process. Therefore, this study proposes utilization of reactive transport modeling to replicate the pore-scale chemo-mechanical reactions and transport processes occurring in silicate-rich shale rocks in the presence of $CO_2$ saturated brine under high pressure and high temperature. For this study, Crunch Tope has been adopted to simulate a one-dimensional reactive transport model of a Permian rock specimen exposed to the acidic brine at a temperature of 100 °C and pressure of 12.40 MPa (1800 psi) for a period of 14 and 28 days. The results demonstrated significant dissolution followed by precipitation of quartz rich phases, precipitation and swelling of clay rich phases, and dissolution of feldspar rich phases closer to the acidic brine-rock interface. Moreover, porosity against reaction depth curve showed nearly 1.00% mineral precipitation occur at 14 and 28 days, which is insufficient to completely fill the pore spaces.

## 1. INTRODUCTION

The subsurface has enormous potential to extract and store energy and may act as a great energy waste repository. However, extensive knowledge of the subsurface characterization and multi-physical processes involved in energy extraction and storage methods is required to successfully implement them. For this reason, four energy technologies such as nuclear waste disposal, energy geo-structures, hydrocarbon and geothermal reservoirs, and geological $CO_2$ storage are widely being investigated to understand the geomechanics involved in the energy storage or extraction processes and to overcome any associated issues (Houhou & Laloui, 2022). These studies assist in analyzing the stability of subsurface or geostructures when exposed to harsh environmental conditions which are typically encountered in applications such as enhanced oil recovery, $CO_2$ sequestration, and geothermal energy extraction. One of the suitable and appropriate geo-structural sites for energy storage like $CO_2$ storage is considered to be shale rocks. Due to their abundant nature, potential sealing capability, intricate porous structure, and lower permeability (Sobhbidari & Hu, 2021; Wu et al., 2020). Still, before adopting these rocks as a potential storage site, a comprehensive understanding of the chemo-mechanical alterations and transport processes occurring in these rocks when subjected to harsh environmental conditions needs to be investigated.

Research has shown that shale rocks develop micro and macro cracks with high or low porosity zones when exposed to $CO_2$-saturated brine. This results in the creation of potential leakage pathways for the stored energy or concentrated damage zones inside the rock (Prakash et al., 2022). Although conventional models assume uniform rock-fluid interactions, studies have shown the evolution of reaction zones with significant mechanical property alterations (Aman et al., 2018; Prakash et al., 2024). The reason for such behavior is attributed to the coupled and dynamic nature of rock-fluid interaction (Prakash et al., 2024). Experimental studies on rocks subjected to $CO_2$-saturated brine have also demonstrated the transformation of feldspar into different mineral groups like quartz, kaolinite, gibbsite, and illite (clay) depending on the physical and chemical conditions (Yuan et al., 2019). Moreover, chemical reactions such as precipitation and dissolution of carbonate (Clark & Vanorio, 2016; Prakash et al., 2019; Vialle & Vanorio, 2011), silicate, and phyllosilicate minerals (Prakash et al., 2024; Prakash & Abedi, 2023), and iron oxidation into ferromagnesian silicate and sulfide minerals (Ulmer-Scholle et al., 2014) are also observed. These chemical reactions may result in microcrack formation, silicate and carbonate cement alterations, and short or long-term debonding (Prakash et al., 2024). Furthermore, a study by Prakash et al. (2024) investigated the chemo-mechanical alterations occurring in silicate-rich shale rocks on exposure to $CO_2$ or $N_2$-rich brine under high temperature (100 °C) and pressure (12.4 MPa) conditions. The experimental testing was conducted for two different durations, 14 and 28 days. The ionic strength of the brine solution was maintained at 1M using NaCl. The results of $CO_2$ exposure showed the initial dissolution of quartz and clay-rich phases followed by their precipitation. The reaction depths at 14 and 28 days were about 1100 μm and 1500 μm, respectively. Reduction of indentation modulus was also observed in the clay and quartz-rich phases, with the quartz-rich phase reducing more than 50% of the initial modulus value at 28 days of reaction time. Additionally, a weaker clay-quartz interface was seen at distances greater than 5 mm away from the reacted surface, due to the potential microcracking of the rock sample induced by the swelling observed in clay particles (Prakash et al., 2024).

While current literature provides some experimental research studies emphasizing the significance of investigating the chemo-mechanical reactions occurring in heterogeneous rocks when exposed to harsh environmental conditions such as $CO_2$-saturated brine, a research gap exists on the pore-scale effect of $CO_2$-saturated brine exposure on these rocks (Khan et al., 2024). Typically, investigation of these heterogeneous rocks (like shale rocks) requires static and dynamic macro-scaled experimentation of drill cores or core plugs, which are quite limited and expensive due to their difficulty in recovering and preserving procedures. Also, due to the heterogeneity of rocks, experimental testing of multiple specimens would be required for adequate understanding and deduction of the effects of $CO_2$-saturated brine, which further increases the cost. Hence, a suitable, efficient, and appropriate alternative is to perform geochemical and reactive transport modeling simulation. These simulation techniques are capable of investigating complex geochemical reactions, transport processes, coupling of geochemical reactions, and fracture permeability (Asadi et al., 2024) in heterogeneous porous rocks with ease, higher efficiency, lower cost, and time.

Presently, some simulation studies (Adila et al., 2023; Asadi et al., 2024; Berrezueta et al., 2023; Fatah et al., 2022; Khan et al., 2024; Liu et al., 2012; Murugesu, 2024) exist that focus on researching the geothermal reactions and transport processes occurring in heterogeneous rocks when exposed to harsh environmental conditions. One such study was conducted by Berrezueta et al. (2023), who adopted CrunchFlow for the geochemical modeling of a gabbro-anthrosite rock specimen. The rock specimen was exposed to two different types of $CO_2$-saturated seawater conditions. The simulation was conducted in two stages under realistic temperature and pressure conditions: Stage I was exposure of the specimen to $CO_2$ supersaturated seawater and Stage II was exposure of the specimen to $CO_2$ subsaturated seawater. The authors compared their simulated results against their experimental findings for validation of the results. The results showed higher iron, magnesium, and calcium concentrations during dissolution in Stage I and carbonation in Stage II (Berrezueta et al., 2023; Khan et al., 2024). CrunchFlow was also utilized by Asadi et al. (2024) to perform reactive transport modeling of Mancos shale rock when exposed to $CO_2$-saturated brine to investigate its implications on the fracture surface of the rock sample. As expected higher reaction was determined to occur at the regions closer to the inlet and fracture surface of the sample (Asadi et al., 2024). Another study using CrunchFlow includes Murugesu (2024), who studied $CO_2$-saturated brine interactions with Wolfcamp shale rocks at pH levels of 2 and 4. Both 1D and 2D reactive transport simulations were conducted at pH levels of 2 and 4. The results demonstrated the dissolution of multiple minerals such as gypsum, dolomite, halite, and pyrite, which led to higher porosity and fluid diffusivity across the fracture and matrix interface. Subsequently, reprecipitation of pyrite and Iron (III) hydroxide occurred, however, the amount of precipitation was reported to be insufficient to close the fracture, which validated their experimental findings and observations (Murugesu, 2024). Still, based on the literature review, a lack of research studies was found which focused on utilizing reactive transport simulation to study the pore-scale chemo-mechanical properties and transport processes of silicate-rich shale rocks on exposure to $CO_2$-saturated brine under high temperature and pressure conditions.

This study attempts to replicate the pore-scale chemo-mechanical alterations and transport processes occurring in silicate-rich shale rocks when subjected to $CO_2$-saturated brine under high temperature and pressure conditions presented by Prakash et al. (2024), using a reactive transport simulation software. CrunchTope was adopted to simulate a 1D reactive transport model consisting of $CO_2$-saturated brine and rock domains. Similar to Prakash et al.'s (2024) study, a Permian rock specimen was utilized with temperature and pressure conditions of 100 °C and 12.40 MPa (1800 psi), respectively. Also, 1M ionic strength of brine was used with the support of NaCl. In addition, for actual replication of the experimental study, the nucleation of amorphous silicon dioxide, kaolinite, and gibbsite was considered. The experimental findings provided by Prakash et al. (2024) have been used in this study for validation of the simulated results. This paper supports in evaluation of the potential of reactive transport simulation to investigate the pore-scale chemo-mechanical reactions and transport processes occurring in heterogeneous rocks when exposed to harsh environmental conditions encountered during energy storage and extraction methods. Hence, it proposes an efficient and quick approach for the investigation of such studies with lower cost and time, by eliminating the need for experimental testing of heterogeneous rock specimens.

## 2. METHODOLOGY

### 2.1. Material Properties

The rock specimen utilized in this study belongs to the Permian formation. The mineral composition (Prakash et al., 2024) and their respective densities are provided in Table 1 (Nguene, 2019). The Permian rock specimen is mainly composed of quartz, clay (illite), and feldspar with minimal amounts of dolomite, pyrite, siderite, and organic matter. The mineral volume fractions were computed using the attained mass fractions and Eq. (1).

$$\eta_i = (1 - \phi) \frac{w_i}{\sum_{k=1}^{n} w_k \frac{\rho_i}{\rho_k}} \tag{1}$$

where $\eta_i$ represents the volume fraction of species i, $\phi$ denotes the porosity, $w_i$ stands for the weight fraction of species i, $w_k$ represents the weight fraction of species k (k = 1,2,…,n), $\rho_i$ is the density of species i and $\rho_k$ denotes the density of species k (k = 1,2,…,n) (Nguene, 2019; Varanasi, n.d.).

Table 1. Mineral composition of the rock specimen(Nguene, 2019; Prakash et al., 2024)

| Mineral Compound | Proportion (%) | Density (g/cm$^3$) |
|---|---|---|
| Quartz | 54.59 | 2.65 |
| Clay (Illite) | 23.31 | 2.75 |
| Feldspar (Albite) | 16.61 | 2.65 |
| Dolomite | 2.54 | 2.86 |
| Pyrite | 1.67 | 4.89 |
| Siderite | 1.27 | 3.96 |
| Total organic content (TOC) | 5.15 | 1.41 |

As mentioned in Prakash et al. (2024), the rock specimen was prepared for indentation and Scanning Electron Microscopy and Energy Dispersive Spectroscopy (SEM-EDS) tests. For these tests, the rock specimen was cut, grinded, and trimmed to a cube with a size of 10 mm. This was followed by coarse polishing of the cubed specimens with 400-grit hard perforated pads (TexMet P, Buehler) and oil-based diamond suspension to prevent any further chemical reactions. Next, the specimens were dry polished using 12-, 9-, 3-, 1- and 0.3-μm aluminum oxide abrasive disks. Also, the specimens were ultrasonicated using n-decane solution (as they are unreactive with the rock) in between the polishing steps. For further details on the specimen preparation steps, the readers are encouraged to refer to (Abedi, Slim, & Ulm, 2016; Abedi, Slim, Hofmann, et al., 2016; Martogi & Abedi, 2020; Sharma et al., 2019).

The prepared specimens were exposed to $CO_2$-saturated brine (1M NaCl synthetic brine) by placing them in a titanium-based batch-type Parr dissolution reactor (with volume of 250 cm$^3$). The temperature and pressure conditions were maintained at 100 °C and 12.40 MPa (1800 psi), respectively. A high liquid-to-solid ratio of 19 was used to attain the appropriate reaction rate. The specimen was exposed for 14 and 28 days (2 and 4 weeks) for appropriate evolution of the reaction zone. For further detailed information about the experimental testing of the rock specimen, the readers can refer to Prakash et al. (2024).

Prakash et al. (2024) also provided the required information about the procedures of different surface tests such as nanoindentation, SEM-EDS, and micro-CT imaging, conducted on the reacted specimen for analyzing the chemo-mechanical properties.

## 2.2. Reactive Transport Simulation Modeling
To replicate and verify the experimental findings, CrunchTope reactive transport code was used to simulate a 1D reactive transport model of the silicate-rich rock specimen when exposed to $CO_2$-saturated brine. It assists in determining the chemo-mechanical reactions and transport processes occurring due to $CO_2$-saturated brine-rock interaction. CrunchTope is an open-source reactive transport code with the ability to perform advective, diffusive, and dispersive transport reactions. This code is also capable of conducting various mixed equilibrium and kinetic reactions with ease (Asadi et al., 2024; Murugesu, 2024; Steefel & Lichtner, 1998).

The governing equation of CrunchTope is shown in Eq. (2) (Li et al., 2017; Nguene, 2019):

$$\frac{\partial(\phi C_i)}{\partial t} = \frac{\partial}{\partial x}\left(D_{ie}\frac{\partial C_i}{\partial x}\right) - \frac{\partial}{\partial x}(\phi u C_i) + \sum_{r=1}^{N} v_{ir}R_{ir} \qquad (2)$$

Where $\phi$ denotes the porosity, $C_i$ stands for the concentration of the species $i$, $t$ represents the time, $x$ stands for the dimension axis, $D_{ie}$ represents the effective diffusion coefficient of species $i$, $u$ represents the average fluid linear velocity, $v_{ir}$ is the reaction stochiometric coefficient of species $i$, and $R_{ir}$ denotes the reaction rate for species $i$ in the $r^{\text{th}}$ reaction.

The effective diffusion coefficient is computed using Eq. (3), which requires the diffusion coefficient, $D_i$, and cementation coefficient, $m$ values.

$$D_{ie} = \phi^m D_i; \ \phi = 1 - \sum_j \phi_{mj} \qquad (3)$$

Where, $\phi_{mj}$ is the volume fraction of the mineral j.

The chemical reactions are performed as per TST rate law using Eq. (4):

$$R = Aka_{H+}^n\left(1 - \frac{IAP}{K_{sp}}\right) \qquad (4)$$

Where, $A$ and $k$ represent the surface area and rate constant, $a_{H+}$ stands for the activity of hydrogen ion ($H+$) with $n$ being the rate dependence on the $H+$ activity, $IAP$ denotes the ion activity product and $K_{sp}$ is the solubility product of the reacting mineral (Li et al., 2017; Nguene, 2019).

A 1D reactive transport model as presented in Fig. 1 was used in this study. The main domains of this model are $CO_2$-saturated brine, and shale rock (Permian). The discretization of the domains adopted in the reactive transport model is as follows: (1) 91 grid cells of $CO_2$-saturated brine with finer cells closer to the shale rock domain, and (2) 70 grid cells of shale rock with finer cells closer to the brine-rock interface and the cell width increases with higher distance away from the acidic brine domain.

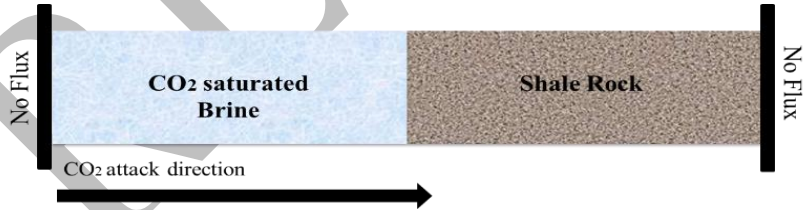

Fig. 1. CrunchTope simulation model framework

For the cementation coefficient of the Permian rock, a value of 2.40 was used in the model, due to the significant tortuosity of the rock specimen. The diffusion coefficient for each species of the rock was set at a value of 1.48 x $10^{-9}$ m²/s for a temperature of 100 °C. Since the diffusivities are in order of $10^{-9}$ m²/s, utilization of a uniform diffusion coefficient is justified and supports in simplifying the modeling process. Additionally, it was observed that inputting the specific diffusion coefficient for each mineral species does not result in a significant increase in the level of accuracy of the results (Li et al., 2017). The porosity and permeability of the reacted Permian rock specimen used was about 5.00% and 1.00 x $10^{-21}$ m² (1 nD), determined based on the experimental testing of the specimen. The model was also updated to include the nucleation of the minerals, such as amorphous silicon dioxide ($SiO_2$(am)), kaolinite, and gibbsite. To simulate the nucleation of these minerals, a thin inert substrate layer was included in the brine domain closer to the brine rock interface. The equations to evaluate the nucleation reaction rate is provided in Eqs. (5-6) (Li et al., 2017):

$$J = J_0 \, exp\left(-\frac{\Delta G^*}{kT}\right) \qquad (5)$$

$$\Delta G^* = \frac{16\pi v^2 a^3}{3k_B^2 T^2 \left[ln\left(\frac{IAP}{K_{sp}}\right)\right]^2} \qquad (6)$$

Where, $J_0$ is the kinetic factor for the rock specimen (= 1.00 x $10^{-8}$ mol/m²/s), $\Delta G^*$ is the nucleation energy barrier, $v$ represents the nucleating phase molecular volume, $a$ is the effective interfacial energy of the rock specimen (= 47 ± 1 mJ/m²), and $k_B$ stands for the Boltzmann's constant (Li et al., 2017).

The mineral solubility products adopted in the simulation model for the Permian rock specimen at a temperature of 100 °C is presented in Table 2.

Table 2. Mineral solubility products for CrunchTope simulation

| Mineral Compound | Mineral solubility product, $\log_{10}(K_{sp})$ at 100 °C |
|---|---|
| Quartz | -3.08 |
| Clay (Illite) | 2.05 |
| Feldspar (Albite) | 0.22 |
| Dolomite | 0.09 |
| Pyrite | -21.23 |
| Siderite | -1.50 |
| SiO₂(am) | -2.18 |

## 3. RESULTS AND DISCUSSION OF RESULTS

### 3.1. Phase Concentration of Minerals at 14 and 28 Days of Reaction Time

Figure 2 presents the concentrations of quartz-rich, clay-rich, and feldspar-rich phases of Permian rock on exposure to $CO_2$-saturated brine for a reaction time of 14 days using CrunchTope simulation and experimentation provided by Prakash et al. (2024). The simulation results demonstrate significant similarity to the experimental findings in the phase concentration trends of quartz-rich, clay-rich, and feldspar-rich phases over the distance from the reacted surface (μm).

Initial dissolution followed by precipitation of quartz-rich phase was noticed with increase in distance from the acidic brine-rock interface to the reaction depth inside the rock. This is shown by the reduction in phase concentration from 54.90% at 30 μm to 48.30% closer to the interface, demonstrating dissolution near the reacted surface. This was followed by precipitation of quartz from 48.30% at 1080 μm to 54.90% at 800 μm of the reacted distance. The dissolution and precipitation trend of the quartz-rich phase observed in the simulated results is similar to the experimental results. Yet, the magnitude of dissolution is not as significant in the simulated results compared to experimental results. As the experimental results in Fig. 2 indicate quartz-rich phase dissolution and degradation from nearly 44.57% at 132 μm to about 13.90% at the brine-rock interface (0 μm), followed by slight precipitation from 36.00% (at 273 μm) to about 44.57%, and this was followed by further dissolution from 61.43% (at 774 μm) to 36.00% inside the rock specimen. The difference in magnitude between the simulated and experimental results is due to the assumption that quartz is a highly stable and insoluble mineral with high resistance against weathering. Hence, CrunchTope simulated results do not seem to demonstrate significant dissolution and degradation in the quartz-rich phase. However, the experimental results have shown significant dissolution and degradation of quartz minerals due to stress corrosion cracking. Additionally, the precipitation in the quartz-rich phase is attributed to the transformation of the feldspar-rich grains in the acidic environment (Prakash et al., 2024).

Precipitation of the clay-rich phase can be noticed, closer to the reacted surface in the simulated results which was also observed in the experimental findings. The phase concentration of the clay-rich phase increased from 19.90% to 27.10% with the increase in distance from the rock specimen to the reacted surface (from 1080 μm to 800 μm). Following this, clay precipitation was maintained at a stable rate of 27.10% till the brine rock interface. Nonetheless, the magnitude of precipitation was not as high as the experimental results. The precipitation observed in experimental results was the result of clay swelling. However, one of the drawbacks of reactive transport simulation software CrunchTope is its inability to simulate the swelling of the clay phase and/or grain detachment (Dávila et al., 2020; Geramian et al., 2016). Moreover, the precipitation occurred till a reacted depth of about 1080 μm in the simulation results, which is significantly close to the experimental reacted depth of 1100 μm. Therefore, the simulated results portray nearly similar reaction depth as the experimental results.

Feldspar-rich phase demonstrated significant dissolution by CrunchTope simulation as well as experimental results. The phase concentration reduced from 14.70% (inside the rock) to 0.00% (at the acidic brine-rock interface). The dissolution of feldspar is due to the transformation of these mineral grains to quartz and clay minerals as a result of the reaction of the rock specimen to $CO_2$-saturated brine. This behavior was also noticed in the experimental results as mentioned in (Prakash et al., 2024). The reaction depth of quartz-rich, clay-rich, and feldspar-rich phases was about 1080 μm, which is close to the experimental observations.

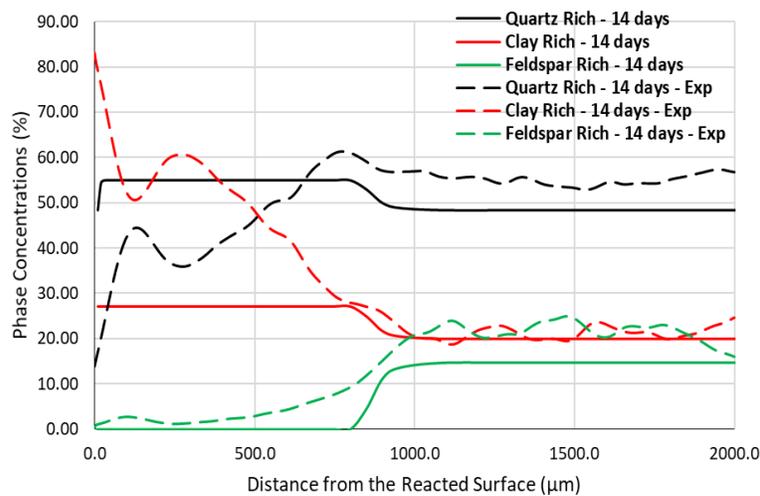

Fig. 2. Quartz-rich, clay-rich, and feldspar-rich phase concentrations of Permian rock at 14 days using CrunchTope simulation and experimental findings from Prakash et al. (2024)

Figure 3 provides the simulated results of phase concentrations over the reacted distance against the experimental findings from (Prakash et al., 2024) at 28 days of reaction time. The phase concentrations of quartz-rich, clay-rich, and feldspar-rich phases demonstrate similar behavior at 14 and 28 days of reaction time (refer to Figs. 2 and 3). So, similar to the reaction of 14 days, 28 days showed dissolution followed by precipitation of the quartz-rich phase, precipitation of the clay-rich phase, and dissolution of the feldspar-rich phase. Even the experimental results showed similar trends at 14 and 28 days but the reaction is found to have reduced in terms of magnitude at longer reaction time. Analogous to Fig. 2, in Fig. 3, the significant amount of dissolution and degradation of the quartz-rich phase seen in the experimental results was not portrayed by the simulation results due to the assumption of quartz being a stable mineral with significant resistance to weathering. In addition, the significant precipitation of the clay-rich phase due to clay swelling was not depicted in the CrunchTope simulation. Nevertheless, feldspar dissolution was depicted by the simulation similar to the experimental findings.

Furthermore, as expected, the reaction depth is higher at 28 days compared to 14 days of reaction time (refer to Figs. 2 and 3). This was also noticed in the experimental results. In the simulated results, the reaction depth increases from 1080 μm at 14 days to about 1480 μm at 28 days of reaction time. For the experimental results, the reaction depth increases from 1100 μm at 14 days to 1500 μm at 28 days of reaction time. This validates that reaction depth is correlated to reaction time, validating Fick's second law. Fick's second law says that the reaction depth is correlated to the square root of time under constant boundary condition concentration. This finding was also deduced by Prakash et al. (2024). Hence, the simulated results are validated by the experimental findings.

Based on the comparison of the simulated and experimental results of quartz-rich, clay-rich, and feldspar-rich phase concentrations against the distance from the reacted surface at 14 and 28 days, one can deduce that CrunchTope is capable of simulating the pore-scale chemo-mechanical reactions and transport processes occurring in silicate-rich shale rocks on exposure to $CO_2$-saturated brine. But, CrunchTope has some limitations such as its inability to consider the swelling of clay minerals and its assumption regarding the stability of quartz minerals against weathering.

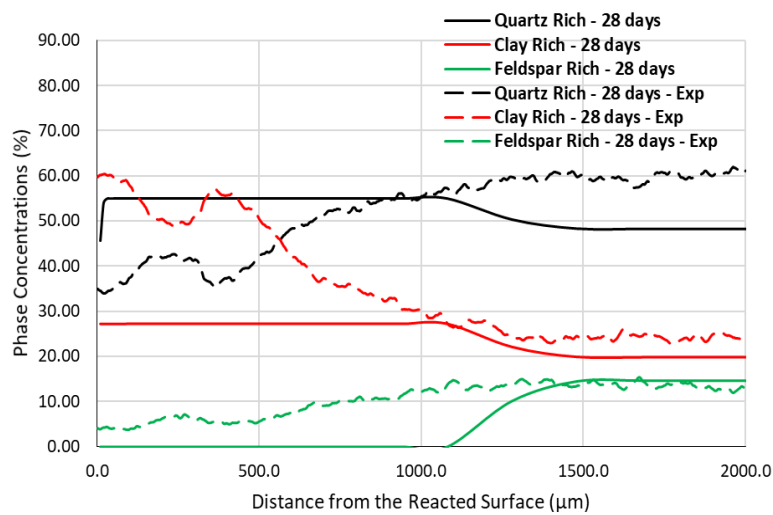

Fig. 3. Quartz-rich, clay-rich, and feldspar-rich phase concentrations of Permian rock at 28 days using CrunchTope simulation and experimental findings from Prakash et al. (2024)

### 3.2. Normalized Phase Concentrations at 14 and 28 Days

The normalized phase concentrations at 14 and 28 days as a function of distance from the reacted surface are presented in Fig. 4. The phase concentrations were normalized with respect to their initial phase concentrations found in the unreacted rock specimen. Parallel to the mineral phase evolutions between 14 and 28 days shown in Figs. 2 and 3, their normalized phase concentration demonstrates similar dissolution/precipitation trends at 14 and 28 days of reaction time. This behavior is also observed in the experimental results shown in Prakash et al. (2024). Furthermore, the magnitude of experimental results is higher for the clay-rich phase and slightly lower in the quartz-rich phase compared to the simulated results. This was the result of the clay swelling and dissolution and degradation of quartz due to stress corrosion cracking experienced during experimentation. However, these cannot be depicted using CrunchTope simulation due to its limitations.

The phase concentrations seem to obtain similar values at 14- and 28 days of reaction time. This behavior can also be observed in the experimental results with similar dissolution/precipitation trends of mineral phases at 14 and 28 days with slight variations in magnitude. Nonetheless, the main difference between 14- and 28-days result is their reaction depth. Longer reaction depth was noticed with an increase in reaction time as per Fick's second law. The reaction depth had increased from 1080 µm at 14 days to 1480 µm at 28 days of reaction time, which is validated by the experimental results. With higher reaction time, there is an increase in the concentration of ions such as $Na^+$, which results in higher precipitation of secondary minerals such as kaolinite (Prakash et al., 2024), leading to clay precipitation. Quartz-rich phase precipitation involves precipitation of amorphous silicon dioxide ($SiO_2$(am)). The evolution of secondary minerals proves the capability of reactive transport code to conduct and simulate the equilibrium speciation and kinetic reactions occurring due to the reaction of the rock minerals with dissolved $CO_2$.

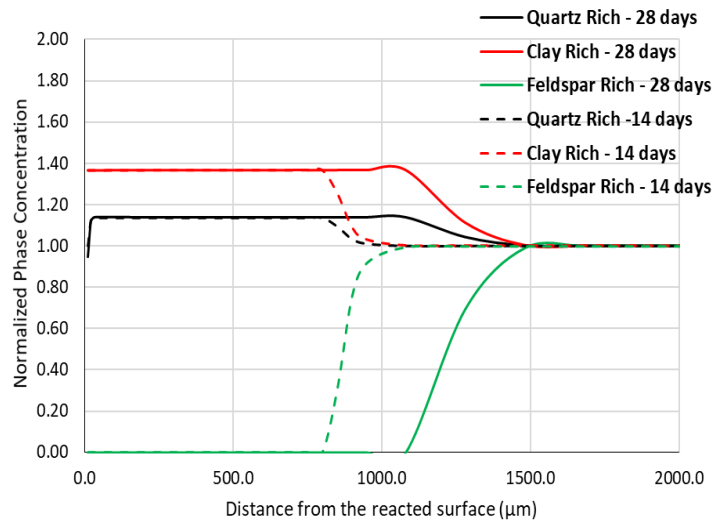

Fig. 4. Normalized phase concentrations of quartz-rich, clay-rich, and feldspar-rich phases of Permian rock at 14 and 28 days using CrunchTope simulation

### 3.3. pH of Brine and Rock Specimen

Figure 5 presents the pH variation with time at the interface of $CO_2$-saturated brine and the rock specimen. pH of the acidic brine increased from 3.38 to 4.78 (at 14 and 28 days). Simultaneously, the Permian rock specimen pH value reduced from 8.48 to 4.78 (at 14 and 28 days). The acidic brine began to lose its acidity on its interaction with the rock specimen. While a simultaneous increase in the acidity of the rock specimen was observed, demonstrating the dissolution of the minerals in the rock specimen due to the reaction with the dissolved $CO_2$ of the acidic brine. This shows that the acidity of the acidic brine reduces on reaction with the rock specimen, which was initially basic in nature.

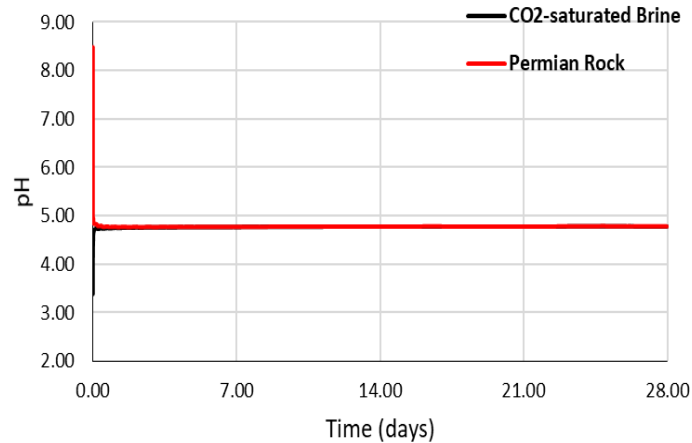

Fig. 5. Variation of pH of $CO_2$-saturated brine and Permian rock over time

### 3.4. Porosity of the Rock Specimen

Figure 6 shows the variation in porosity of the rock specimen with the distance from the reacted surface. The porosity was seen to increase initially, followed by a slight reduction, and succeeded by a further increase in porosity. The increase in porosity demonstrates mineral dissolution whereas, the reduction in porosity represents precipitation of the minerals. Based on the porosity variation, the reaction depth was about 1080 μm and 1480 μm at 14 and 28 days, respectively. These reaction depths are similar to the phase concentration results. Furthermore, Fig. 6 results show that the porosity increases (demonstrating mineral dissolution) with an increase in distance from the rock specimen towards the brine-rock interface. The porosity had increased from 5.00% to 15.30% at 14 days and from 5.00% to 17.90% at 28 days. Hence, there is significant mineral dissolution occurring in the rock specimen, which has been stated in the earlier results. Nevertheless, a slight precipitation of about 0.89% (from 450 μm to 210 μm) and 1.19% (from 500 μm to 270 μm) was noticed in the rock specimen at 14 and 28 days, respectively. However, the precipitation of minerals is found to be significantly low, so complete filling of the pore spaces was not achieved in either of the reaction times. Therefore, this shows that the exposure of silicate-rich shale rocks to $CO_2$-saturated brine under high temperature and pressure conditions does not result in the complete filling of the pores, so further diffusion of the acidic brine inside the rock specimen may occur over time. This observation was also reported by Li et al. (2017) for cement rocks with secondary precipitation of calcium carbonate ($CaCO_3$).

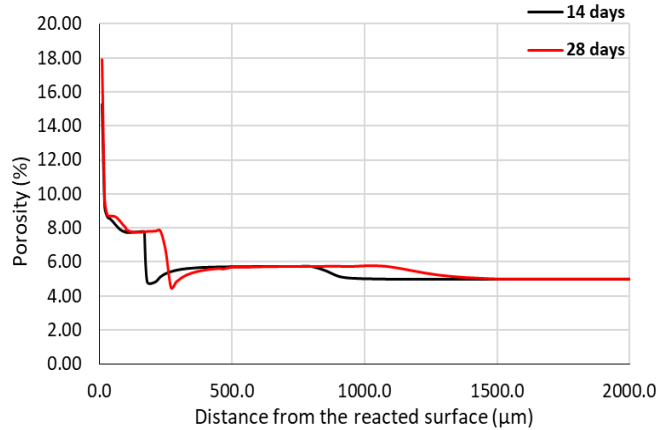

Fig. 6. Porosity variation with respect to the distance from the reacted surface of the rock specimen

## 4. CONCLUSION

This paper utilizes reactive transport simulation to investigate the pore-scale chemo-mechanical reactions and transport processes occurring in silicate-rich shale rocks when exposed to $CO_2$-saturated brine under high temperature and pressure conditions. Permian rock specimen and CrunchTope reactive transport code were used for this study. The simulation is performed for 14 and 28 days of reaction time at a temperature of 100 °C and pressure of 12.40 MPa (1800 psi). The simulated results are validated against the experimental findings provided by Prakash et al. (2024). The following findings were deduced from this study:

- Quartz-rich phase showed initial dissolution, followed by slight precipitation with an increase in distance from the acidic brine-rock interface to deeper depths inside the rock specimen. At 14 days, the phase concentration decreased from 54.90% at 30 μm to 48.30% at the acidic brine-rock interface, indicating mineral dissolution near the reacted surface. This was followed by quartz precipitation from 48.30% at 1080 μm to 54.90% at 800 μm of the reacted distance. Similar dissolution/precipitation trends were observed at 28 days.
- Quartz precipitation included the precipitation of amorphous silicon dioxide ($SiO_2$(am)).
- Phase concentration of the clay-rich phase demonstrates clay precipitation at 14 and 28 days of reaction time, owing to the precipitation of kaolinite mineral. The phase concentration of the clay-rich phase increased from 19.90% to 27.10% with the increase in distance from the rock specimen to the reacted surface.
- Dissolution of feldspar was observed in the simulated results due to feldspar grain transformation to quartz and clay minerals, similar to the experimental results. The phase concentration reduced from 14.70% (inside the rock) to 0.00% (at the acidic brine-rock interface).
- The reaction depth of quartz-rich, clay-rich, and feldspar-rich phases at 14 and 28 days were about 1080 μm and 1480 μm, respectively, which is close to the experimental observations.
- Quartz-rich phase dissolution magnitude was lower compared to the experimental results. As CrunchTope assumes quartz to be a stable mineral with high resistance to weathering and higher insolubility. However, the experimental results showed significant quartz degradation due to stress corrosion cracking.

- Clay-rich phase showed slight precipitation in the simulated results compared to the experimental findings. This was attributed to the fact that CrunchTope was unable to simulate the swelling of clay particles.
- The normalized phase concentrations showed similar behavior as the experimental results but varied in terms of magnitude due to the clay swelling and dissolution and degradation of quartz minerals.
- Reaction of $CO_2$-saturated brine with Permian rock specimen resulted in reducing the acidity of acidic brine. A simultaneous increase in the acidity of the rock specimen was observed demonstrating acidic brine-rock interaction.
- The porosity against the distance for the reacted surface portrayed significant dissolution followed by slight precipitation (about 0.89% at 14 days and 1.19% at 28 days) of minerals. This observation was similar to the phase concentration results. The slight precipitation of minerals was unable to fill the pores of the rock specimen, so further diffusion of the acidic brine may occur inside the rock over time.

Thus, CrunchTope successfully simulated the pore-scale transport processes and chemo-mechanical reactions occurring in the silicate-rich shale rocks when exposed to $CO_2$-saturated brine under high temperature and pressure conditions. The results were validated against their experimental findings. Also, the reactive transport simulation showed precipitation of secondary minerals such as kaolinite and amorphous silicon dioxide ($SiO_2$(am)) which shows its ability to conduct and simulate the required equilibrium reactions and kinetic reactions occurring in the rock specimen due to $CO_2$ dissolution. Hence, based on the findings of this study, reactive transport simulation can be considered as a quick, efficient, and appropriate technique for investigating the pore-scale transport and chemo-mechanical reaction processes of rock specimens when subjected to harsh environmental conditions, with lower cost and time. Also, two key limitations of CrunchTope were determined in this study which include, its inability to depict quartz dissolution and degradation due to stress corrosion cracking and the swelling of clay particles. One of the simulation codes capable of conducting reactive transport modeling studies considering swelling of clay particles is CrunchClay (Tournassat & Steefel, 2019). Still, further research is needed to enhance and update the reactive transport simulation codes with the ability to incorporate key processes such as clay swelling and stress corrosion cracking of quartz.

## ACKNOWLEDGEMENT

The authors would like to gratefully acknowledge the financial support provided by the National Science Foundation (NSF) through Award 2045242, which enabled the completion of this project.